\documentclass[%
reprint,
superscriptaddress,
nofootinbib,
 amsmath,amssymb,
 aps,
 prd,
 twocolumn
]{revtex4-2}

\usepackage{graphicx}% Include figure files
\usepackage{dcolumn}% Align table columns on decimal point
\usepackage{bm}% bold math
\usepackage{xcolor}

\newcommand{\dif}{\,\mathrm{d}}

\newcommand{\ta}{\tilde{a}}
\newcommand{\tA}{\tilde{A}}
\newcommand{\tomega}{\tilde{\omega}}

\begin{document}

%\preprint{APS/123-QED}

\title{Quasinormal modes of gravitational perturbation for uniformly accelerated black holes}

\author{Tan Chen}
\email{chentan@itp.ac.cn}
\affiliation{CAS Key Laboratory of Theoretical Physics, Institute of Theoretical Physics, Chinese Academy of Sciences, Beijing 100190, China}
\affiliation{School of Physical Sciences, University of Chinese Academy of Sciences, Beijing 100049, China}

\author{Rong-Gen Cai}
\email{cairg@itp.ac.cn}
\affiliation{School of Physical Science and Technology, Ningbo University, Ningbo 315211, China}
\affiliation{CAS Key Laboratory of Theoretical Physics, Institute of Theoretical Physics, Chinese Academy of Sciences, Beijing 100190, China}

\affiliation{School of Fundamental Physics and Mathematical Sciences, Hangzhou Institute for Advanced Study, University of Chinese Academy of Sciences, Hangzhou 310024, China}

\author{Bin Hu}
\email{bhu@bnu.edu.cn}
\affiliation{Department of Astronomy, Beijing Normal University, Beijing 100875, China}

\date{\today}

\begin{abstract}
%We calculate the gravitational quasinormal modes of the rotating accelerating black holes described by the spinning C-metric. 
%We prove that the master equations for the massless perturbations of the spinning C-metric with any spins can be transformed into Heun's equations. 
%Two methods are used to calculate the quasinormal frequency and the angular separation constant. 
%The transformation of the master equations makes these two methods quicker and more precise. 
%We identify three distinct sets of modes: the photon sphere modes, the near-extremal modes, and the acceleration modes. 
%No unstable fundamental quasinormal mode is found in this paper. 
%We analyze the influences of spin and acceleration on the quasinormal frequencies in detail. 
%The acceleration parameter has very distinct influences on the quasinormal modes with different azimuthal numbers, especially for the $m = 1$ modes. 
We first show that the master equations for massless perturbations of accelerating rotating black holes can be transformed into the Heun's equation. 
The quasinormal modes of the black holes can be easily calculated in the framework of the Heun's equation. 
We identify three modes for the tensor perturbations: the photon sphere modes, which reduce to the quasinormal modes of Kerr black holes when the acceleration parameter vanishes; the near-extremal modes, which branch from the first set and become dominant when the spin is near extremal; and the acceleration modes, which are closely related to the acceleration horizon. 
We calculate the frequency spectrum of the QNMs in various spin and acceleration parameters.
We choose an angular boundary condition that keeps the angular function regular at $ \theta = 0 $ and $\pi$, which is consistent with the boundary condition of the Kerr black hole. 
%We find that the $m_0 = 1$ modes have anomalous behavior with such a boundary condition.
The conical singularity caused by the acceleration influences this boundary condition.
We find that the $m_0 = 1$ modes have an anomalous behavior at particular accelerations.

%\begin{description}
%\end{description}
\end{abstract}

\maketitle

%\tableofcontents

\section{\label{sec:intro} Introduction}
Black holes are among the strangest and most fascinating objects in the universe. 
Their existence has been confirmed by astrophysical observation~\cite{Remillard:2006fc, EventHorizonTelescope:2019dse}. 
In recent years, the detection of gravitational waves has made it possible to explore the strong regime of gravity around black holes in a completely new way~\cite{LIGOScientific:2016aoc, LIGOScientific:2016vlm}. 
The gravitational waves produced by black hole binaries have three stages -- the inspiral, merger, and ringdown. 
The ringdown phase starts when the black holes approach each other within the photon sphere. 
In this stage, the black hole (BH) remnant can be described by a perturbed state of a black hole solution. 
The emitted gravitational waves have characteristic decay time scales and are well described by the quasinormal modes (QNMs)~\cite{LIGOScientific:2016lio, Franchini:2023eda}. 
The QNM spectrum can be used to perform ``black hole spectroscopy''\cite{Dreyer:2003bv}. 
According to the no-hair theorem~\cite{Israel:1967wq, Carter:1971zc, Robinson:1975bv}, the spectrum acts as a fingerprint of the system and only depends on the parameters of the background black hole.

Astrophysical black holes are naturally neutral and rotating, which can be well described by the Kerr metric. Binary rotating black holes have so far been the primary sources of gravitational waves. 
Most research about gravitational wave sources is limited to non-accelerating black holes, while astrophysical processes can produce accelerating black holes. In particular, the emission of gravitational waves tends to have a preferred direction, which results in the black hole remnant having a recoil acceleration after the merger. 
This process is called the black hole superkick~\cite{Merritt:2004xa, PhysRevD.77.124047, Gerosa:2016vip, CalderonBustillo:2018zuq}. 
Besides, cosmic strings, which are line-like topological defects emerging during first-order phase transitions~\cite{Kibble:1976sj, Vilenkin:1984ib}, can break or fray to produce a pair of accelerating black holes~\cite{Hawking:1995zn, Eardley:1995au}. 
Analysis of accelerating black holes could produce more information about the early universe and astrophysical environments.

A natural choice to describe accelerating rotating black holes is the spinning C-metric. 
This metric describes two causally separated black holes accelerating away from each other by a force corresponding to the tension of a cosmic string~\cite{Griffiths_2006, Griffiths:2005se}. 
With an appropriate choice of coordinates, this metric can be used to cover only one of the black holes. The spinning C-metric has two conical singularities because of the acceleration. 
It has been shown that the conical singularity can be removed by adding an external electromagnetic field~\cite{Hong:2004dm}. 
The C-metric has been used to describe the accelerating supermassive black holes \cite{PhysRevD.107.044031, PhysRevLett.129.031102}, where the gravitational lensing effect of the C-metric is studied. 
In addition, the scalar QNMs of the charged C-metric have been examined in \cite{Destounis:2020pjk, Destounis:2020yav}; the scalar QNMs of the spinning C-metric have been analyzed in \cite{Xiong:2023usm}. 
Although these two metrics are completely different, their QNM spectra share many similarities. 
They both have three distinct sets of modes: the photon sphere modes, the acceleration modes, and the near-extremal modes. 
Their acceleration modes are all closely related to the acceleration horizon. 
However, the gravitational QNMs of the C-metric have not been considered yet. 

In this work we are going to study the gravitational quasinormal modes of the spinning C-metric in detail. 
To obtain the QNMs, we first show that the perturbation equations for the spinning C-metric with any spin weights can be transformed into the Heun's equation~\cite{Heun_1888, Ronveaux1995HeunsDE}.
%, which is easier to deal with. 
Then we develop two numerical methods to calculate the QNMs. 
Using the Heun's equation form of the perturbation equations makes the numerical computation quicker and more precise.
Moreover, these two methods are not limited to the gravitational case. 
They can be used to compute the QNMs of the spinning C-metric with any spin weights. 
We also identify three distinct sets of quasinormal modes for the gravitational perturbation.
Following the convention of the scalar QNMs, we call them the photon sphere modes, near-extremal modes, and acceleration modes. 
The spin and acceleration parameters' influences on the QNMs are carefully analyzed, including the near-extremal cases.
We choose an angular boundary condition that keeps the angular function regular at $\theta = 0$ and $\pi$,
which reduces to the boundary condition of the Kerr black hole when the acceleration parameter vanishes.
We find that the $ s=-2, l=2, m_0=1 $ modes have an anomalous behavior with this boundary condition. 

This paper is organized as follows. 
We review the spinning C-metric and rederive the master equations for the massless perturbations in Section \ref{sec:pert}. 
In Section \ref{sec:method}, we first prove that the master equations can be transformed into the Heun's equation. 
Then we introduce the two numerical methods we used to calculate the QNMs. 
We show the numerical results of the gravitational QNMs in Section \ref{sec:results}. 
Section \ref{sec:conclu} is devoted to conclusion and discussion.
We set $ c = G = 1 $ throughout the paper for brevity.

\section{\label{sec:pert} Perturbations of the spinning C-metric}

\subsection{\label{sec:bkmetric} Background spacetime}
The spinning C-metric belongs to the general Plebański-Demiański family~\cite{Hong:2004dm, Podolsky:2021zwr}. 
It describes a pair of causally separated BHs that accelerate uniformly in opposite directions~\cite{Griffiths_2006}.
Using the Boyer-Lindquist-type coordinates, the spinning C-metric can be expressed as~\cite{Griffiths:2005se}
\begin{eqnarray}\label{eq:cmetric}
		\dif s^2 &=& \frac{1}{\Omega^2} \bigg \{ -\frac{1}{\Sigma} (Q-a^2 P \sin^2 \theta) \dif t^2 \nonumber \\
		& &+ \frac{2a \sin^2 \theta}{\Sigma} [Q - P(r^2 + a^2)] \dif t \dif \varphi  \nonumber \\
		& &+ \frac{\Sigma}{Q} \dif r^2 + \frac{\Sigma}{P} \dif \theta^2 \nonumber \\
		& &+ \frac{\sin^2 \theta}{\Sigma} [ P(r^2+a^2)^2-a^2 Q \sin^2 \theta ] \dif \varphi^2 \bigg \},
\end{eqnarray}
where the functions $ \Omega, \Sigma, P $ and $Q$ are given  by
\begin{eqnarray}
		\Omega &=& 1- Ar \cos \theta, \qquad \Sigma = r^2 +a^2 \cos^2 \theta,  \nonumber \\
		P &=& 1-2AM \cos \theta + a^2 A^2 \cos^2 \theta, \nonumber \\
		Q &=& (1-A^2r^2)(r^2 -2Mr+a^2).
\end{eqnarray}
The parameters $M$, $A$, and $a$ stand for the BH mass, acceleration, and spin, respectively. 
This metric reduces to the C-metric for $a = 0$, to the Kerr metric for $A= 0$, and to the Rindler metric when $M = a =0$~\cite{Griffiths_2006, Siahaan:2018wvh}. 
The spinning C-metric has a Kerr-like ring singularity at $r = 0, \theta = \pi/2 $. There are three null hypersurfaces at		
\begin{equation}
	r_\pm = M \pm \sqrt{M^2 - a^2}, \quad r_A = \frac{1}{A},
\end{equation}		
which are called the event horizon, Cauchy horizon, and acceleration horizon, respectively. 
We only consider the area $ r_+ < r < r_A$, which implies $ Q > 0 $, $ P > 0 $ for $ \theta \in [0,\pi] $.

There exist conical singularities at the axis $\theta = 0$ and $ \theta = \pi $, corresponding to the existence of deficit angles. 
These singularities cannot be removed simultaneously, unless some
external fields are introduced~\cite{Hong:2004dm, Astorino:2016xiy}. 
Here we specify $ \varphi \in [0,2\pi/P(\pi)) $ to remove the conical singularity at $\theta = \pi$. 
The metric can then be interpreted as a Kerr-like BH being accelerated along the axis $\theta = 0$ by the action of a force that corresponds to the tension of a cosmic string~\cite{Griffiths_2006, Podolsky:2021zwr}.

In the following discussions, it is convenient to introduce the surface gravities $ \kappa $ and angular velocities $ \omega_H $ on various horizons. 
They are defined by
\begin{eqnarray}
	\kappa(r_i) &=& \frac{Q_{,r}}{2(r^2 + a^2)}\bigg|_{r = r_i}, \\
	\omega_H(r_i) &=& \frac{a}{r_i^2 + a^2}.
\end{eqnarray}
The surface gravities are unique up to a normalization of the associated Killing vector. 

\subsection{\label{sec:perturbME} Master equations for the massless perturbations}
The spinning C-metric is a Petrov type D vacuum metric\cite{10.1093/oso/9780198503705.001.0001}. The perturbation equations for it can be obtained by using the Teukolsky equation in the context of the Newman-Penrose formalism~\cite{Teukolsky:1973ha,Newman:1961qr}. 
Surprisingly, the perturbation equations of the spinning C-metric can also be separated, as proved in \cite{Bini:2008mzd}. 
Here we redrive the equations in the signature $ (-, +, +, +) $ and correct some typos in \cite{Bini:2008mzd}. 
For the spinning C-metric, we adopt the following null tetrad
\begin{eqnarray}
		l^\mu &=& \Omega^2 \left( \frac{r^2+a^2}{Q}, 1, 0, \frac{a}{Q} \right), \nonumber \\
		n^\mu &=& \frac{1}{2 \Sigma} (r^2 +a^2, -Q, 0, a), \nonumber \\
		m^\mu &=& \frac{\Omega}{\sqrt{2 P} (r+i a \cos \theta)} \left( i a \sin \theta, 0, P, \frac{i}{\sin \theta }\right), \nonumber \\
		\bar{m}^\mu &=& \frac{\Omega}{\sqrt{2 P} (r- ia \cos \theta)} \left( -i a \sin \theta, 0, P, -\frac{i}{\sin \theta }\right).
	\label{eq:tetrad}
\end{eqnarray}
They satisfy 
\begin{equation}
	l^\mu n_\mu = -1, \quad m^\mu \bar{m}_\mu =1,
\end{equation}
while all other scalar products are zero.
The choice of the null tetrad has 6 degrees of freedom. We first use 4 degrees of freedom to choose a null tetrad satifying 
\begin{eqnarray}
		\Psi_0 &=& \Psi_1 = \Psi_3 = \Psi_4 = 0, \nonumber \\
		\kappa &=& \sigma = \nu = \lambda  = 0,
\end{eqnarray}
so that the Teukolsky equation holds. Here $\Psi_0, \Psi_1, \Psi_3 $ and $\Psi_4$ are the Weyl scalars. $ \kappa, \sigma, \nu$ and  $\lambda$ are the spin coefficients\footnote{Only here $\kappa, \epsilon, \delta, \gamma, \alpha, \beta$ stand for the spin coefficients.}. These are the scalar quantities used in the Newman-Penrose formalism~\cite{Newman:1961qr} and their definitions are shown in Appendix~\ref{app:NPquantity}. 
The remaining 2 degrees of freedom are used to set the spin coefficient $\epsilon = 0$.
% Using the Newman-Penrose formalism~\cite{Newman:1961qr}, we have\footnote{Only here $\kappa, \epsilon, \delta, \gamma, \alpha, \beta$ stand for the spin coefficients.}
% \begin{eqnarray}
% 		\Psi_0 &=& \Psi_1 = \Psi_3 = \Psi_4 = 0, \nonumber \\
% 		\kappa &=& \sigma = \nu = \lambda = \epsilon = 0.
% \end{eqnarray}
Therefore, using the null tetrad from Eq.(\ref{eq:tetrad}), the only nonzero Weyl scalar is
\begin{equation}
	\Psi_2 = (1+iaA) M \rho_0^3,
\end{equation}
where 
\begin{equation}
	\rho_0 = -\frac{\Omega}{(r-ia \cos \theta)}.
\end{equation}
The nonzero spin coefficients are
\begin{eqnarray}
		\rho &=& \rho_0 (1 - i a A \cos^2 \theta), \nonumber \\
		\alpha &=& \pi - \beta^* + \frac{\sqrt{2P} Ar \sin \theta}{(r-ia\sin \theta)}, \nonumber \\
		\mu &=& \frac{Q}{2 \Omega^2 \Sigma} \rho,  \nonumber \\
		\pi &=& - \frac{\sqrt{P} (r^2 A -ia)\sin \theta }{\sqrt{2} (r-ia\cos \theta)^2}, \nonumber \\
		\tau &=& \frac{\sqrt{P} (r^2 A - ia) \sin \theta }{\sqrt{2} \Sigma}, \nonumber \\
		\gamma &=& \mu + \frac{Q_{,r}\Omega + 4 Q A \cos \theta}{4\Omega \Sigma}, \nonumber \\
		\beta &=& - \frac{\sqrt{P}}{2\sqrt{2}} \cot \theta \left( \frac{\rho^*}{\Omega} + A\cos \theta \right)  \nonumber \\
		& & + \frac{\Omega}{2\sqrt{2}} \frac{(\sqrt{P})_{,\theta}}{(r+ia \cos \theta)}.
\end{eqnarray}

The Teukolsky equations for massless perturbations of any spin weights $ s $ can be cast into a compact form~\cite{Vagenas:2020bys, Arbey:2021jif}
\begin{eqnarray}\label{eq:teu1}
		\{ &&[D- \rho^* + \epsilon^* + \epsilon - 2s(\rho+\epsilon)] (\Delta+\mu-2s\gamma) \nonumber \\
		&&- [\delta + \pi^* - \alpha^* + \beta - 2s(\tau+\beta)] (\delta^* + \pi -2s \alpha)  \nonumber \\
		&&- 2(s-1)(s-1/2) \Psi_2	\} \Psi = 0
\end{eqnarray}
for $ s=1/2, 1, 3/2, 2 $ and 
\begin{eqnarray}\label{eq:teu2}
		\{ &&[ \Delta - \gamma^* + \mu^* - \gamma - 2s (\gamma+\mu) ] (D-\rho -2s\epsilon) \nonumber \\
		&&- [ \delta^* - \tau^* + \beta^* - \alpha - 2s(\alpha+\pi) ] (\delta - \tau -2s \beta )  \nonumber \\
		&&-2 (s+1)(s+1/2)\Psi_2  \} \Psi = 0	
\end{eqnarray}
for $ s=-1/2, -1, -3/2, -2 $. $ D, \Delta, \delta, \bar{\delta} $ are directional derivatives defined by
\begin{equation}
	D \equiv l^i \nabla_i, \quad \Delta \equiv n^i \nabla_i, \quad \delta \equiv m^i \nabla_i , \quad \bar{\delta} \equiv \bar{m}^i \nabla_i.
\end{equation}
$ \Psi $ represents the perturbation field with different spin weights. Its definition is shown in Table \ref{tab:1}, where the symbols for different fields are the same as those in \cite{Teukolsky:1973ha}, except that $ H_0 , H_1 $ stand for the components $ H_{000}, H_{111} $ of the Rarita–Schwinger field~\cite{10.1063/1.528409}. 
%Only here $\Delta$ stands for the directional derivative.

%\begin{table}[b]
%	\caption{\label{tab:1}  Spin-weight $s$ and the field quantity $\Psi$ for Eq.~(\ref{eq:maseq}).	}
%	\begin{ruledtabular}
%		\begin{tabular}{cccccccccc}
%			s & 0 &1/2 &-1/2 &1  &-1 &-3/2 &-3/2 &2  &-2\\
%			\colrule
%			$ \Psi $& $ \Phi $ & $\chi_0 $ & $ \chi_1 $ & $\phi_0 $ & $ \phi_2$ & $H_0$ & $ H_1 $ & $\psi_0$ &  $ \psi_4$\\
%		\end{tabular}
%	\end{ruledtabular}
%\end{table}

\begin{table}[b]
	\caption{\label{tab:1}  Spin-weight $s$, field quantity $\Psi$ in Eq.~(\ref{eq:teu1}-\ref{eq:teu2}), and field quantity $\psi$ in Eq.~(\ref{eq:maseq}).	}
	\begin{ruledtabular}
		\begin{tabular}{ccc}
			s & $\Psi$ in Eq.~(\ref{eq:teu1}-\ref{eq:teu2}) & $\psi$ in Eq.~(\ref{eq:maseq})\\
			\colrule
			0 & $ \Phi $ & $ \Phi $\\
			1/2 & $\chi_0 $ & $\chi_0 $ \\
			-1/2 & $ \chi_1 $ & $\rho_0^{-1} \chi_1 $\\
			 1  & $\phi_0 $ & $\phi_0 $\\
			-1  & $ \phi_2$ & $\rho_0^{-2} \phi_2$ \\
			3/2 & $H_0$     & $H_0$ \\
			-3/2 & $ H_1 $  & $\rho_0^{-3} H_1 $ \\
			 2  & $\psi_0$ & $\psi_0$ \\
			-2  & $ \psi_4$ & $ \rho_0^{-4} \psi_4$\\
		\end{tabular}
	\end{ruledtabular}
\end{table}

%\begin{table}[htb]
%	\centering
%	\begin{tabular}{|c|c|c|c|c|c|c|c|c|c|}
%		\hline
%		s & 0 &1/2 &-1/2 &1  &-1 &-3/2 &-3/2 &2  &-2\\
%		\hline
%		$ \psi $&  & $\chi_0 $ & $\rho_0^{-1} \chi_1 $ & $\phi_0 $ & $\rho_0^{-2} \phi_2$ & $H_0$ & $\rho_0^{-3} H_1 $ & $\psi_0$ &  $ \rho_0^{-4} \psi_4$\\
%		\hline
%	\end{tabular}
%	\caption{\label{tab:2} Spin-weight $s$ and the field quantity $\psi$ for equation \ref{eq:maseq}.}
%\end{table}
To obtain the perturbation equations for the spinning C-metric, we choose the null tetrad from Eq.~(\ref{eq:tetrad}) and substitute Eqs.~(8-11) into the Teukolsky equations. The resulted equations can be combined into a master equation
\begin{equation}\label{eq:maseq}
	\left[(\nabla^\mu - s \Gamma^\mu) (\nabla_\mu - s \Gamma_\mu) + 4 s^2 \Psi_2 \right] \psi = 0,
\end{equation}
where we have defined a connection vector
\begin{eqnarray}
		\Gamma^t &=& \frac{\Omega^2}{\Sigma} \bigg \{ \frac{1}{Q} \left[ M (A^2 r^4 + a^2) \nonumber \right.\\
		& &\left. + r (1+a^2 A^2) (r^2 - 3Mr + a^2)  \right] \nonumber \\
		& &+ \frac{ia}{P} \left[ (1+a^2 A^2) \cos \theta - AM (1+\cos^2 \theta) \right] \bigg\}, \nonumber \\
		\Gamma^r &=& - \frac{\Omega}{\Sigma} \left( \frac{1}{2} \Omega Q_{, r} + 2 A \cos \theta Q \right), \nonumber \\
		\Gamma^\theta &=& \frac{2A\Omega P r \sin \theta}{\Sigma}, \nonumber \\
		\Gamma^\varphi &=& - \frac{\Omega^2}{\Sigma} \bigg[ \frac{a Q _{,r}}{2 Q} + i \frac{\cos \theta (2P-1)}{P\sin^2 \theta} \nonumber \\
		& &+ i \frac{AM(\cos^2 \theta + 1) -A^2 a^2 \cos \theta}{P\sin^2 \theta}\bigg].
\end{eqnarray}	
$ \psi $ in Eq.~(\ref{eq:maseq}) is a redefined field quantity.
Its definition is also shown in Table \ref{tab:1}.

The master equation can be separated by writing~\cite{Bini:2008mzd}
\begin{equation}
	\psi(t,r,\theta,\varphi) = \Omega^{1+2s} e^{-i\omega t} e^{im\varphi} R(r) S(\theta),
\end{equation}
where $ \omega $ is the quasinormal frequency and $ m $ is the azimuthal number. 
Since the exponent $ m \varphi $ should have the period $2\pi$ and $ \varphi $ has been redefined to remove the conical singularity, $ m $ must be of the form $ m= m_0 P(\pi) $, where $ m_0 $ is an integer.

The separated radial equation is
\begin{equation}\label{eq:radeq}
	Q^{-s} \frac{\dif}{\dif r} \left( Q^{s+1} \frac{\dif R(r)}{\dif r} \right) + V_{\text{rad}} (r) R(r) = 0,
\end{equation}
with
\begin{eqnarray}
		V_{\text{rad}}(r) &=& -2rA^2 (r-M) (1+s)(1+2s) \nonumber \\ 
		& &+ \frac{ [(r^2+a^2) \omega - am]^2 }{Q} \nonumber \\
		& &- 2is \bigg [- \frac{am Q_{,r}}{2Q} + \frac{\omega M (r^2-a^2)}{r^2 -2 Mr + a^2} \nonumber \\
		& &- \frac{\omega r (1+a^2 A^2)}{1-A^2 r^2} \bigg ]+A_{lm}.
\end{eqnarray}
$ A_{lm} $ is the separation constant. 
For $ A = 0$, the radial equation is equivalent to the perturbation equation of the Kerr black hole~\cite{Teukolsky:1973ha}. The only differences are the definitions of the separation constants and they are related by
\begin{equation}
	A_{lm} = - A_{lm}^\text{Kerr} - a^2 \omega^2 + 2am\omega,
	\label{eq:alimit}
\end{equation}
where $A_{lm}^\text{Kerr}$ is the separation constant in \cite{Teukolsky:1973ha}.

%The angular part is
%\begin{equation}\label{eq:angeq}
%	\frac{1}{\sin \theta} \frac{\dif}{\dif \theta} \left( \sin \theta \frac{\dif Y(\theta)}{\dif \theta} \right) + V_{\text{ang}} (\theta) Y(\theta) = 0,
%\end{equation}
%where 
%\begin{equation}
%	\begin{aligned}
	%		V_\text{ang} (\theta) &= \frac{1-A_{lm}+s(1-a^2A^2)}{P} + \frac{1}{P^2} \left\{ - \frac{[w \cos \theta-s(1+a^2A^2)]^2}{\sin^2 \theta} \right. \\
	%		&-(z+w-4sAM)^2 + [z \cos \theta -s(1+a^2A^2)]^2-(1-AM\cos\theta)^2 \\
	%		&\left. -a^2A^2+A^2M^2+4sa^2A^2\cos\theta(2sAM-w)  \right\},
	%	\end{aligned}			
%\end{equation}
%with
%\begin{equation}
%	Y(\theta) = \sqrt{P} S(\theta) , z=a\omega+sAM , w = -m+2sAM. 
%\end{equation}
Defining $ u = \cos \theta$, the separated angular equation can be expressed as
\begin{equation}\label{eq:angeq}
	\frac{\dif}{\dif u} [P (1 - u^2)  \frac{\dif S(u)}{\dif u}] + V_{\text{ang}} (u) S(u) = 0,
\end{equation}
where 
\begin{eqnarray}
		V_{\text{ang}}(u) &=& -\frac{[ (1-u^2) \tomega +(1+\ta^2)su -\tA s (1+u^2) - m ]^2}{P (1-u^2) }  \nonumber \\
		& & - \frac{4 m s u}{1-u^2} - u P_{,u} + s(1-\ta^2) -A_{lm}.	
\end{eqnarray}
The definitions of $ \ta, \tA, \tomega $ are
\begin{equation}
	\ta = A a, \quad \tA = A M, \quad \tomega = a \omega. 
\end{equation}
In order to compute the QNM frequencies, we need to solve the eigenvalue problems of Eqs.~(\ref{eq:radeq}) and (\ref{eq:angeq}) with appropriate boundary conditions. 
The physically motivated boundary conditions for the radial part are
\begin{equation}
	R(r) \sim \left \{
	\begin{array}{ll}
		( r - r_+ )^{-s-i\frac{\omega - m \omega_H(r_+)}{2 \kappa(r_+)}}, \quad &r \rightarrow r_+, \\
		(r_A - r)^{i \frac{ \omega - m \omega_H(r_A)  }{2\kappa (r_A)}}, \quad &r \rightarrow r_A.
	\end{array}
	\right.
	\label{eq:rbd}
\end{equation}
These conditions correspond to that the waves propagate only inward at the event horizon and only outward at the acceleration horizon. 
We also require the solution to be finite at the interval boundaries of $\theta$, which gives the boundary conditions
\begin{equation}
	S(\theta) \sim \left \{
	\begin{array}{ll}
		(1-\cos \theta)^{\frac{1}{2} \left| s + \frac{m}{P(0)} \right|  },   \quad &\theta \rightarrow 0, \\
		(1+\cos \theta)^{\frac{1}{2} \left| -s + \frac{m}{P(\pi)} \right|  },  \quad &\theta \rightarrow \pi.
	\end{array}
	\right.
	\label{eq:abd}
\end{equation}
The conical singularities cause the boundary conditions to be different from the Kerr case~\cite{Leaver:1985ax}.
The additional coefficient $1/P$ of $m$ represents the deficit angle.
After the redefinition of $\varphi$, the boundary conditions become
\begin{equation}
    S(\theta) \sim \left \{
	\begin{array}{ll}
		(1-\cos \theta)^{\frac{1}{2} \left| s + m_0 \frac{P(\pi)}{P(0)} \right|  },   \quad &\theta \rightarrow 0, \\
		(1+\cos \theta)^{\frac{1}{2} \left| -s + m_0 \right|  },  \quad &\theta \rightarrow \pi.
	\end{array}
	\right.
	\label{eq:abdm}
\end{equation}
We can see that the conical singularity at $\theta = \pi$ is removed.

\section{\label{sec:method} Solution to the perturbation equations}
In this section, we show that Eqs.~(\ref{eq:radeq}) and (\ref{eq:angeq}) can be transformed
into the Heun’s equation. 
Then we use two numerical methods to obtain the QNMs. 
The first is the continued fraction method. 
The second is the shooting method. Both methods rely on transforming the perturbation equations into the Heun's equation and can be used to compute QNMs with any spin weights.

\subsection{Transformation of the perturbation equations into the Heun’s equation}
Heun's equation is a second-order differential equation with four singular points~\cite{Heun_1888, Ronveaux1995HeunsDE}. 
It can be expressed as
\begin{eqnarray}
	\frac{\dif^2 w}{\dif z^2} &+& \left( \frac{\gamma}{z} + \frac{\delta}{z-1} + \frac{\epsilon}{z-z_0} \right) \frac{\dif w}{\dif z} \nonumber \\
	&+& \frac{\alpha \beta z - q }{z (z-1) (z-z_0)} w = 0,
	\label{eq:heun}
\end{eqnarray}
with $ \gamma + \delta + \epsilon = \alpha + \beta + 1 $.
This equation has Frobenius solutions in the neighborhood of a singular point. 
The recursion relation between the expansion coefficients can be written in an analytic three-term form. 
The perturbation equations for Kerr-de Sitter black hole have been transformed into the Heun's equation by Suzuki, Takasugi, and Umetsu in~\cite{Suzuki:1998vy}. 
Based on that, the numerical calculation to obtain the QNMs becomes more rapid and precise~\cite{Yoshida:2010zzb, Hatsuda:2020sbn}. 
Here we show that the perturbation equations for the spinning C-metric can also be transformed into the Heun's equation.

\subsubsection{\label{sec:angheun} Angular Perturbation equation}
The angular perturbation equation Eq.~(\ref{eq:angeq}) has 5 five regular singularities at $ u= -1, 1, u_+, u_- $ and $\infty$, with $ u_\pm = \frac{1}{\tA \pm \sqrt{\tA^2 - \ta^2}} $.
By using the new variable 
\begin{equation}
	z = \frac{(u + 1) (1-u_-)}{2(u-u_-)},
	\label{eq:zvar}
\end{equation}
the angular equation becomes 
\begin{eqnarray}
	\frac{\dif^2 S(z)}{\dif z^2} &+& \left(\frac{1}{z} + \frac{1}{z-1} + \frac{1}{z-z_+} - \frac{2}{z-z_\infty} \right) \frac{\dif S(z)}{\dif z } \nonumber \\
	&-& \frac{V_{\text{ang}}(z) S(z)}{2 \ta^2(u_+ - u_-) z(z-1)(z-z_+) } = 0.
\end{eqnarray}
This equation has regular singularities at $ z = 0, 1, z_+, \infty $ and $ z_\infty $, with $ z_+ = \frac{(u_+ + 1) (1-u_-)}{2(u_+ - u_-)} $ and $ z_\infty = \frac{1-u_-}{2} $. 
The above equation can be further simplified to the form
\begin{eqnarray}
		\frac{\dif^2 S(z)}{\dif z^2} &+& \left(\frac{1}{z} + \frac{1}{z-1} + \frac{1}{z-z_+} - \frac{2}{z-z_\infty} \right) \frac{\dif S(z)}{\dif z } \nonumber \\
		&+&\bigg[ -\frac{A_1^2}{z^2} -\frac{A_2^2}{(z-1)^2} -\frac{A_3^2}{(z-z_+)^2}  \nonumber \\
		&+&\frac{2}{(z-z_\infty)^2} + \frac{A'_1}{z} + \frac{A'_2}{z-1} + \frac{A'_3}{z-z_+}  \nonumber \\
		&-& \frac{\frac{1}{z_\infty} + \frac{1}{z_\infty -1 } + \frac{1}{z_\infty - z_+}}{z-z_\infty} \bigg] S(z),
		\label{eq:asimp}
\end{eqnarray}
where $ A'_1, A'_2 $ and $ A'_3 $ are given in Appendix~\ref{app:coefficients}.
The regular singularity at $ z = z_\infty $ can be factored out by the following transformation
\begin{equation}
	S(z) = z^{A_1}  (z - 1)^{A_2} (z - z_+)^{A_3} (z- z_\infty) f(z),
	\label{eq:atrans}
\end{equation}
where
\begin{eqnarray}
		A_1 &=& \frac{1}{2} \left| -s + \frac{m}{P(\pi)}  \right| , \nonumber \\
		A_2 &=& \frac{1}{2} \left| s + \frac{m}{P(0)}  \right|, \nonumber \\
		A_3 &=& \frac{1}{2} \bigg\{ s +  \frac{m[\ta^4 + \ta^2 - 2\tA (\tA + \sqrt{\tA^2-\ta^2})]}{\sqrt{\tA^2-\ta^2} P(0)P(\pi)} \nonumber \\
		& &-\frac{\tomega}{\sqrt{\tA^2-\ta^2}}   \bigg\}.
\end{eqnarray}
%\begin{equation}
%	A_3 = \frac{1}{2} \left[ s - \frac{\tomega}{\sqrt{ \bar{M}^2 - \ba^2 }} + \frac{ m (\ba^4 - A^2 r_+^2 )}{\sqrt{\bM^2 -\ba^2} P(0) P(\pi)} \right].
%\end{equation}
Then, $ f(z) $ satisfies the Heun's equation
\begin{eqnarray}
	\frac{\dif^2 f(z)}{\dif z^2} &+& \left( \frac{\gamma_a}{z} + \frac{\delta_a}{z-1} + \frac{\epsilon_a}{z-z_+} \right) \frac{\dif f(z)}{\dif z} \nonumber \\
	&+& \frac{\alpha_a  \beta_a z - q_a }{z (z-1) (z-z_+)} f(z) = 0,
	\label{eq:aheun}
\end{eqnarray}
with
\begin{eqnarray}
		\gamma_a &=& 2A_1 + 1, \nonumber \\
		\delta_a &=& 2A_2 + 1, \nonumber \\
		\epsilon_a &=& 2A_3 + 1, \nonumber \\
		\alpha_a &=& 1 + A_1 + A_2 - \frac{\tomega}{\sqrt{\tA^2 - \ta^2}}  \nonumber \\
		& &+ \frac{m(\ta^4 + \ta^2 -2\tA^2)}{\sqrt{\tA^2 - \ta^2}P(0)P(\pi)} , \nonumber \\
		\beta_a &=& 1 + s + A_1 + A_2 - \frac{2 A m M}{P(0)P(\pi)} , \nonumber \\
		q_a &=&  z_+ (2 A_1 A_2+A_1+A_2)+2 A_1 A_3 + A_1 \nonumber \\
		& & +A_3 + A_1^2 +\frac{1}{2}- \frac{1}{2 \ta^2 (u_+ - u_-)} \bigg[ A_{lm} \nonumber \\
		& & -(1-\ta^2 -m)s -\frac{1}{2} (1-\ta^2) (s + \frac{m}{P(\pi)})^2 \nonumber \\
		& & + \frac{m^2}{P(\pi)} - 2(s+\frac{m}{P(\pi)})\tomega \bigg].
\end{eqnarray}
%\begin{equation}
%	\begin{aligned}
	%		q_a &= A_1^2 + A_1 A_2 + 2A_1 A_3 + \frac{1}{2} (1 + A_2 + 2 A_3 + 3 A_1) \\
	%		&+\frac{1}{8\sqrt{\bM^2 - \ba^2}} \bigg \{  (1-\ba^2)[ (1 + 2A_1)(1 + 2A_2) - (1+s)^2] -4 s \tomega + 2 A_{lm}  \\
	%		&+ \frac{m^2 - 4 m \tomega + 4 s m (\ba^2 + \bM)}{P(\pi)} + \frac{2m^2 (\ba^2 + \bM)}{P(\pi)^2} \bigg \}.
	%	\end{aligned}
%\end{equation}
We can see that $ \gamma_a + \delta_a + \epsilon_a = \alpha_a + \beta_a + 1 $.

The boundary condition Eq.~(\ref{eq:abd}) is transformed into
\begin{equation}
	f(z) \sim \left \{
	\begin{array}{ll}
		1, \quad &z \rightarrow 0, \\
		1, \quad &z \rightarrow 1.
	\end{array}
	\right.
	\label{eq:ahbd}
\end{equation}

\subsubsection{Radial Perturbation equation}
The radial perturbation equation has 5 five regular singularities at $ r= r_+, r_A, -r_A, r_- $, and $\infty$, respectively. By using the new variable
\begin{equation}
	x = \frac{r - r_+}{r - r_-} \frac{r_A - r_-}{r_A - r_+},
	\label{eq:xvar}
\end{equation}
Eq.~(\ref{eq:radeq}) is transformed into an equation with regular singularities at $ 0, 1, x_-, \infty $ and $ x_\infty $, with
\begin{equation}
	x_- = \frac{-r_A - r_+}{-r_A - r_-} \frac{r_A -r_-}{r_A - r_+}, \quad x_{\infty} = \frac{r_A - r_-}{r_A - r_+}.
\end{equation}
The regular singularity at $ x = x_\infty $ can be factored out by the transformation
\begin{equation}
	R(x) = x^{B_1}  (x - 1)^{B_2} (x - x_-)^{B_3} (x- x_\infty)^{2s+1} g(x),
	\label{eq:rtrans}
\end{equation}
where
\begin{eqnarray}
		B_1 &=& -s-i\frac{\omega - m \omega_H(r_+)}{2 \kappa(r_+)}, \nonumber \\
		B_2 &=& i \frac{ \omega - m \omega_H(r_A)  }{2\kappa (r_A)}, \nonumber \\
		B_3 &=& i \frac{ \omega - m \omega_H(-r_A)  }{2\kappa (-r_A)}.
\end{eqnarray}
Then, $ g(x) $ satisfies the Heun's equation
\begin{eqnarray}
	\frac{\dif^2 g(x)}{\dif x^2} &+& \left( \frac{\gamma_r}{x} + \frac{\delta_r}{x-1} + \frac{\epsilon_r}{x-x_-} \right) \frac{\dif g(x)}{\dif x} \nonumber \\
	 &+& \frac{\alpha_r  \beta_r x - q_r }{x (x-1) (x-x_-)} g(x) = 0,
	 \label{eq:rheun}
\end{eqnarray}
with
\begin{eqnarray}
		\gamma_r &=& 2B_1 + 1 + s, \nonumber \\
		\delta_r &=& 2B_2 + 1 + s, \nonumber \\
		\epsilon_r &=& 2B_3 + 1 + s, \nonumber \\
		\alpha_r &=& 2B_1 + 3s + 1, \nonumber \\
		\beta_r &=& 2B_2 + 2B_3 + 1 , \nonumber \\
		q_r &=& x_-[(B_1+B_2)(1+s)+2B_1 B_2] + (B_1 + B_3)(1+s)\nonumber \\
		& & + 2 B_1 B_3 + \frac{1}{(1- A r_+) (1+ A r_-)} \bigg[-A_{lm} \nonumber \\
		& &+ (1 - A^2 a^2 )(1+s)(1+2s) -i \frac{2 a s (m - 2 a \omega)}{r_+ - r_-} \nonumber \\
		& &- \frac{8 i m B_3 \omega_H(r_A) r_-}{(r_+ - r_-) \kappa(r_A) } + \frac{16 A^2 M a B_2 B_3 \omega_H(r_+)}{\kappa(r_+)} \nonumber \\
		& &+ \frac{16 M B_3 (B_1 + s)\omega_H(r_A)}{a \kappa(r_A)}  \bigg].
\end{eqnarray}
We can see that $ \gamma_r + \delta_r + \epsilon_r = \alpha_r + \beta_r + 1 $.

The boundary condition Eq.(\ref{eq:rbd}) becomes
\begin{equation}
	g(x) \sim \left \{
	\begin{array}{ll}
		1, \quad &x \rightarrow 0, \\
		1, \quad &x \rightarrow 1.
	\end{array}
	\right.
	\label{eq:rhbd}
\end{equation}

\subsection{Continued fraction method}
The continued fraction method (Leaver method)~\cite{Leaver:1985ax} is one of the most precise methods to compute QNMs~\cite{Franchini:2023eda}. 
It can be used to find high-accuracy QNMs up to a moderate range of overtones $n$~\cite{Berti:2009kk, Konoplya:2011qq, Yoshida:2010zzb}. 
%The main idea of this method is to construct a Frobenius series expansion that satisfies the boundary conditions. Then the recurrence relation for the coefficients of the series of the solution can be written in terms of continued fraction. 
After we get the Heun's equation form of the perturbation equations, we can directly use the continued fraction method to compute the QNMs. 
Consider the angular perturbation equation, a Frobenius solution satisfying the boundary condition (\ref{eq:abd}) can be expressed as
\begin{equation}
		S(z) = z^{A_1}  (z - 1)^{A_2} (z - z_+)^{A_3} (z- z_\infty) \sum_{n=0}^{\infty} a_n z^n,
		\label{eq:aseries}
\end{equation}
where $ z $ is defined by Eq.~(\ref{eq:zvar}).
According to the last section, the series $ f(z) = \sum_{n=0}^{\infty} a_n z^n $ should satisfy the Heun's equation Eq.~(\ref{eq:aheun}). 
Then the expansion coefficients $a_n $ are defined by a three-term recurrence relation
\begin{eqnarray}
		&&c^1_0 a_{1} +c^2_0 a_0 = 0, \nonumber \\
		&&c^1_n a_{n+1} +c^2_n a_n + c^3_n a_{n-1} =0, \quad (n\ge 1)
		\label{eq:arecursion}
\end{eqnarray}
%The coefficients $ c^1_n, c^2_n, c^3_n $ are
where
\begin{eqnarray}
	c^1_n &=& z_+ (n+1)(n+\gamma_a), \nonumber \\
	c^2_n &=& -(z_+ + 1)n^2 - [z_+ (\gamma_a+\delta_a-1) \nonumber \\
	& &+\gamma_a+\epsilon_a -1]n - q_a, \nonumber \\
	c^3_n &=& n^2 + (\gamma_a + \delta_a + \epsilon_a -3) (n-1) \nonumber \\
	& &+ \alpha_a \beta_a -1.
\end{eqnarray}
Here we choose the initial coefficient to be $ a_0 = 1 $.

The infinite series in Eq.~(\ref{eq:aseries}) converges only if the corresponding solutions $ a_n $ for the recurrence relation (\ref{eq:arecursion}) is minimal. 
This condition is equivalent to the condition in terms of continued fraction~\cite{Leaver:1985ax, Yoshida:2010zzb}, given by
\begin{eqnarray}
	0 &=& c^2_0 - \frac{c^1_0 c^3_1}{c^2_1 - \frac{c^1_1 c^3_2}{c^2_2 - (c^1_2 c^3_3 / c^2_3 - \cdots)}}, \nonumber \\
	&\equiv& c^2_0 - \frac{c^1_0 c^3_1}{c^2_1 - } \frac{c^1_1 c^3_2}{c^2_2 - } \frac{c^1_2 c^3_3}{c^2_3 - } \cdots.
	\label{eq:acfm}
\end{eqnarray}
Given specific $M, a, A, s, m$, Eq.~(\ref{eq:acfm}) is an equation for $ \omega $ and $ A_{lm} $. 
The angular number $l$ can be specified by using Eq.~(\ref{eq:alimit}) and the continuity of $ A_{lm} $ as a function of $A$.
Similarly, the solution for the radial perturbation equation can be expressed as
\begin{equation}
	R(x) = x^{B_1}  (x - 1)^{B_2} (x - x_-)^{B_3} (x- x_\infty)^{2s+1} \sum_{n=0}^{\infty} b_n x^n,
\end{equation}
where $ x $ is defined by Eq.~(\ref{eq:xvar}).
The expansion coefficients $b_n $ are defined by the three-term recurrence relation
\begin{eqnarray}
	&&d^1_0 b_{1} +d^2_0 b_0 = 0, \nonumber \\
	&&d^1_n b_{n+1} +d^2_n b_n + d^3_n b_{n-1} =0. \quad (n\ge 1)
	\label{eq:rrecursion}
\end{eqnarray}
We also choose the initial value to be $ b_0 = 1 $. 
The coefficients $ d^1_n, d^2_n, d^3_n $ are
\begin{eqnarray}
	d^1_n &=& x_- (n+1)(n+\gamma_r), \nonumber \\
	d^2_n &=& -(x_- + 1)n^2 - [x_- (\gamma_r+\delta_r-1) \nonumber \\
	& &+\gamma_r+\epsilon_r -1]n - q_r, \nonumber \\
	d^3_n &=& n^2 + (\gamma_r + \delta_r + \epsilon_r -3) (n-1) \nonumber \\
	& &+ \alpha_r \beta_r -1.
\end{eqnarray}
The convergence condition is given by
\begin{equation}
	0 = d^2_0 - \frac{d^1_0 d^3_1}{d^2_1 - } \frac{d^1_1 d^3_2}{d^2_2 - } \frac{d^1_2 d^3_3}{d^2_3 - } \cdots.
	\label{eq:rcfm}
\end{equation}
We can get $ \omega $ and $ A_{lm} $ by solving Eqs.~(\ref{eq:acfm}) and (\ref{eq:rcfm}) simultaneously. 
Moreover, the continued fraction method is very powerful at computing overtones. 
The $n$-th overtone is usually found to be the most stable numerical root of the $n$-th inversion of the radial continued fraction~\cite{Leaver:1985ax, Berti:2009kk}.

\begin{figure*}[htbp]
	\centering
	\includegraphics[width=0.45\textwidth]{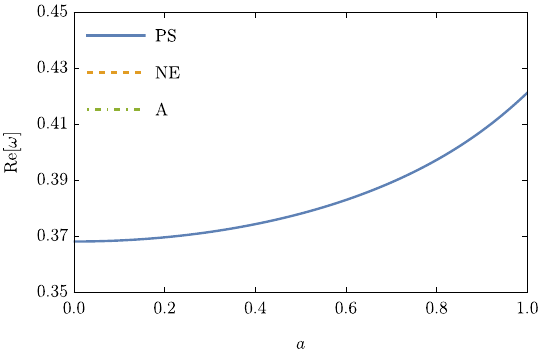}
	\includegraphics[width=0.46\textwidth]{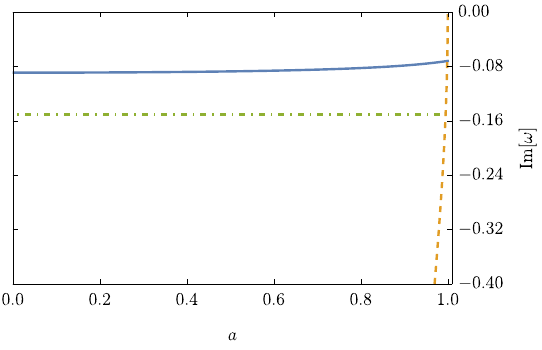}
	\includegraphics[width=0.435\textwidth]{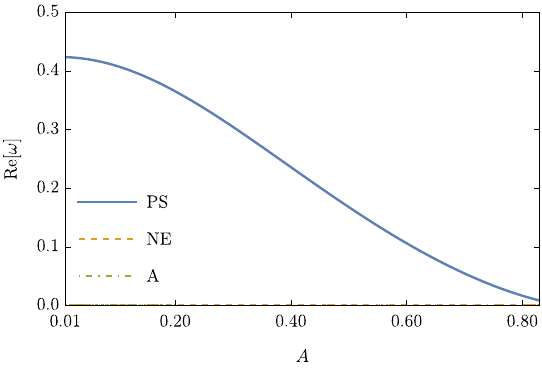}
	\includegraphics[width=0.47\textwidth]{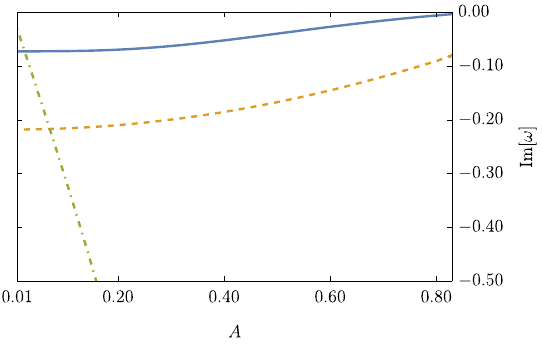}
	\caption{
            Real (left) and imaginary (right) parts of the fundamental modes for all three sets with $ s=-2$, $l=2$, $m_0 = 0$. 
            The upper panels show the frequencies as a function of $ a $ with fixed $ A = 0.05 $. 
            The real parts of the NE mode and A mode vanish and we do not show them in the upper left panel.
            We only show part of the NE mode because its imaginary part decreases drastically when $a$ decreases.
            The bottom panels show the frequencies as a function of $ A $ with fixed $ a = 0.99 $.  
            We can only reach $ A =0.83 $ during our numerical computation.
            The extremal value of $ A $ is $ 0.876$. 
            Similarly, the imaginary part of the A mode decreases drastically with $A$ increasing and we only show part of it.
	}
	\label{fig:allsets}
\end{figure*}

%\begin{figure*}[htbp]
%	\centering
%	\includegraphics[width=0.45\textwidth]{s2psa02r.pdf}
%	\includegraphics[width=0.45\textwidth]{s2psa02i.pdf}
%	\caption{
%		Real (left) and imaginary (right) parts of $ n=0 $ PS modes for $ s=-2$, $l=2$, $m_0 = 2, 1, 0, -1, -2$ with $a=0.2$. The extreme value of $ A $ is $0.505$. 
%	}
%	\label{fig:psa02}
%\end{figure*}

\subsection{Shooting method via the Heun function}
The shooting method is a well-known numerical approach to solving differential equations. 
It was first used by Chandrasekhar and Detweiler to compute the black hole QNMs~\cite{Chandrasekhar:1975zza}. 
The idea is that we integrate the perturbation equation from the boundaries with an initial value for the QNM frequency and match the numerical solutions at an intermediate point. 
If the QNM frequency $ \omega $ is an eigenvalue, the solutions are linearly dependent and the Wronskian of the two solutions should vanish. The QNM frequencies are the corresponding roots. 

To use this method, we usually need to construct the series approximation at boundaries to get the initial values and integrate the equation to get the numerical solution. But for the Heun's equation, this process can be largely simplified. 
The solution to the Heun's equation is called the Heun function, which has been analyzed thoroughly. 
Hatsuda pointed out that we could directly use the Heun function to compute the QNMs~\cite{Hatsuda:2020sbn}. 
We follow this idea and use \textit{Mathematica}'s built-in Heun function to compute the QNM frequencies.

The Heun function $H \ell (z_0, q; \alpha, \beta, \gamma, \delta; z) $ denotes the solution of Eq.~(\ref{eq:heun}) that corresponds to the exponent $ 0 $ and value $1$ at $ z=0 $.
Consider the angular perturbation equation, the solution satisfying the boundary condition Eq.~(\ref{eq:ahbd}) at $ z=0 $ is
\begin{equation}
	f_\text{in}(z) = H \ell (z_+ , q_a; \alpha_a, \beta_a, \gamma_a, \delta_a; z),
\end{equation}
where $ H\ell (z_+ , q_a; \alpha_a, \beta_a, \gamma_a, \delta_a; z) $ is the Heun function and  $ H\ell (z_+ , q_a; \alpha_a, \beta_a, \gamma_a, \delta_a; 0) = 1 $. 
Similarly, the solution satisfying the boundary condition at $ z=1 $ is
\begin{equation}
	f_\text{out}(z) = H\ell (1 - z_+, \alpha_a \beta_a - q_a; \alpha_a, \beta_a, \delta_a, \gamma_a; 1 - z).
\end{equation}
At a midpoint $z_m$, we construct the Wronskian determinant
\begin{equation}
	\text{Det} = \left | \begin{matrix}
		f_{\text{in}}(z_m) & \partial_z f_{\text{in}}(z_m)   \\
		f_{\text{out}}(z_m) &\partial_z f_{\text{out}}(z_m)   \\
	\end{matrix} \right | .
\end{equation}

For the radial equation, we can also construct the Wronskian determinant by the same process. 
$\omega$ and $ A_{lm} $ can be obtained by setting these two determinants to zero.

\begin{figure*}[htbp]
	\centering
	\includegraphics[width=1\textwidth]{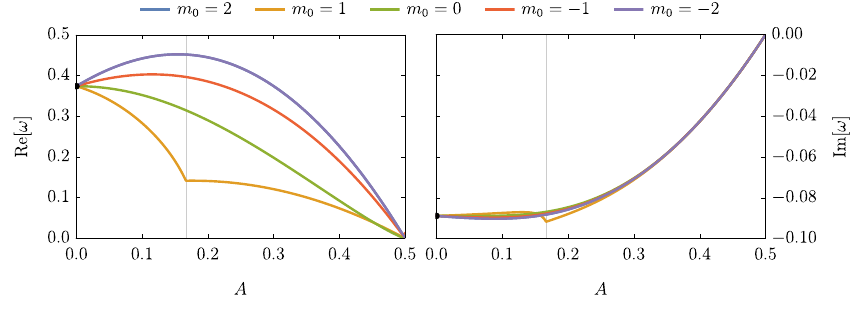}
	\caption{
            Real (left) and imaginary (right) parts of $ n=0 $ PS modes for $ s=-2$, $l=2$, $m_0 = 2, 1, 0, -1, -2$ with $a=0$. 
            The black point represents the fundamental mode of the Schwarzschild black hole. 
            The extremal value of $ A $ is $0.5$. 
            The vertical line represents the value of $ A $ that satisfies $ \frac{P(\pi)}{P(0)} = 2 $.
	}
	\label{fig:psa0}
\end{figure*}

\begin{figure*}[htbp]
	\centering
	\includegraphics[width=1\textwidth]{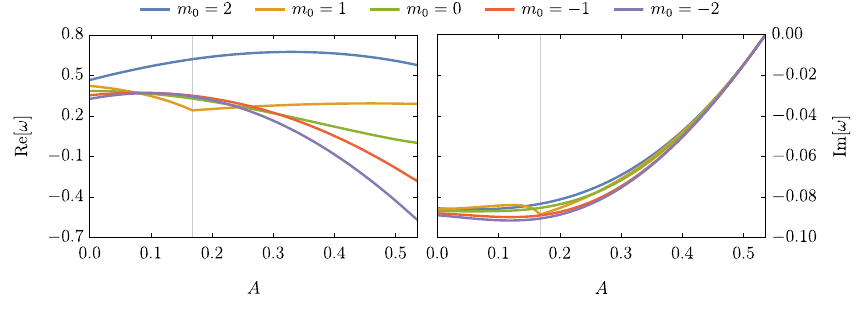}
	\caption{
            Real (left) and imaginary (right) parts of $ n=0 $ PS modes for $ s=-2$, $l=2$, $m_0 = 2, 1, 0, -1, -2$ with $a=0.5$. 
            The extremal value of $ A $ is $0.535$. 
            The vertical line represents the value of $ A $ that satisfies $ \frac{P(\pi)}{P(0)} = 2 $.
	}
	\label{fig:psa05}
\end{figure*}

\section{\label{sec:results} Numerical results}
In this section, we explore the QNM spectrum of the spinning C-metric. 
We mainly focus on gravitational, quadrupolar QNMs ($s = -2, l = 2$). 
These modes may have astrophysical relevance, because the $l = 2, m = 2$ mode dominates the ringdown stage for Kerr black holes~\cite{Buonanno:2006ui, PhysRevD.76.064034}. 
The two numerical methods mentioned above are used to compute the results. 
We justify our results by a direct comparison of the calculated QNM frequencies and separation constants from these methods. 
We further cross-check the results with~\cite{Destounis:2020pjk, Xiong:2023usm} for $ s=0 $ and with~\cite{Leaver:1985ax, Berti:2009kk} in the limit $ A \rightarrow 0 $. 
These results are shown in Appendix \ref{sec:commethod}.  
During the numerical computation, the mass $ M $ is fixed to $1$ and all the physical quantities are expressed as dimensionless forms. 
No unstable fundamental quasinormal mode is found in this paper.

For scalar perturbations ($ s=0 $), it is obvious that $ -\omega^* $ and $ A_{lm}^* $ are the frequency and separation constant for the mode with the azimuthal number $ -m $ if $ \omega $ and $ A_{lm} $ are the eigenvalues for the mode with $m$ from a symmetry of Eqs.~(\ref{eq:radeq}), (\ref{eq:angeq}) and the boundary conditions (\ref{eq:rbd}), (\ref{eq:abd}). 
From our numerical results, we find that this symmetry remains for QNMs with any spin weights $s$ except the $ m_0 = \pm 1 $ cases.
The quasinormal modes of the Kerr black hole possess the same symmetry for all modes~\cite{Leaver:1985ax}. 
For simplicity, we only consider the modes whose real part of $ \omega $ is positive in the limit of $ A\rightarrow 0 $ or $ a \rightarrow 0 $ henceforth. 

\begin{figure*}[htbp]
	\centering
	\includegraphics[width=1\textwidth]{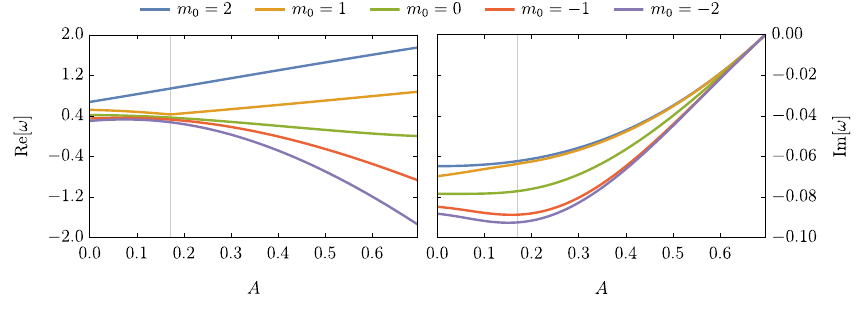}
	\caption{
            Real (left) and imaginary (right) parts of $ n=0 $ PS modes for $ s=-2$, $l=2$, $m_0 = 2, 1, 0, -1, -2$ with $a=0.9$. 
            The extremal value of $ A $ is $0.696$. 
            The vertical line represents the value of $ A $ that satisfies $ \frac{P(\pi)}{P(0)} = 2 $.
	}
	\label{fig:psa09}
\end{figure*}

\begin{figure*}[htbp]
	\centering
	\includegraphics[width=1\textwidth]{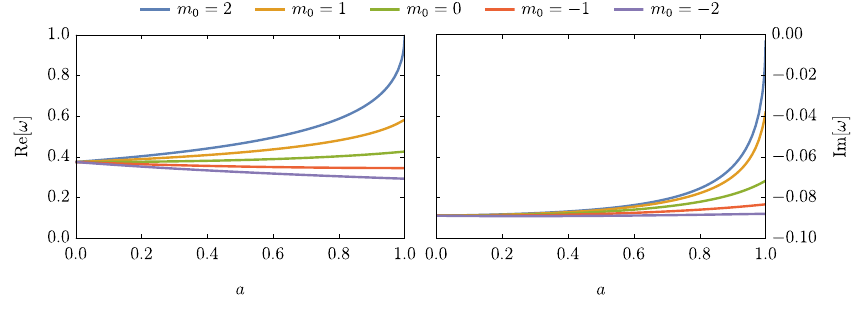}
	\caption{
            Real (left) and imaginary (right) parts of $ n=0 $ PS modes for $ s=-2$, $l=2$, $m_0 = 2, 1, 0, -1, -2$ with $A=10^{-4}$.
	}
	\label{fig:psac0}
\end{figure*}

The numerical results in \cite{Xiong:2023usm} show that there exist three distinct sets of scalar QNMs in the spinning C-metric. 
We also find three sets of QNMs for the gravitational perturbations, but they have different properties from the scalar QNMs.
Following their convention, we label these three sets of QNMs as photon sphere (PS) modes, near-extremal (NE) modes, and acceleration (A) modes. 
Here we mainly present the results for the fundamental modes. 
The fundamental QNM is the one that has the largest imaginary part and thus decays most slowly. 
This mode is usually labeled by the overtone number $ n = 0 $. 
The modes with smaller imaginary parts are labeled as $ n = 1, 2, \cdots $ in sequence. 
In Fig.~\ref{fig:allsets}, we show all these three sets of QNMs for $ s=-2, l=2, m_0 = 0 $ as a function of the parameter $a$ or $ A $.  
The blue solid line is the photon sphere mode, which is the dominant mode among most of the parameter space. 
The yellow dashed line represents the near-extremal mode. 
We can see that this mode dominates the spectrum in the extreme limit $ a \rightarrow 1 $. 
The green dot-dashed line is the acceleration mode.
It is the most distinguishable one. 
Their imaginary parts are almost linearly dependent on the acceleration parameter $ A $ and they decay most slowly when $ A $ is small. 
In the following sections, we introduce these three sets of QNMs in detail.

%\subsection{Spectrum}
%Show the combined plot of three families. As shown in Figs.~\ref{fig:psa0}-\ref{fig:psa09}, all the frequencies of the quasinormal modes of the Spinning C-metric that are calculated in this paper have negative imaginary parts and unstable mode has not been found.
%
%It is important to note that all QNMs computed in the subextremal parameter space of the charged C−metric have Im($ \omega $) < 0, which indicates that such spacetime is modally stable against neutral massless scalar perturbations. 

\begin{figure*}[htbp]
	\centering
	\includegraphics[width=1\textwidth]{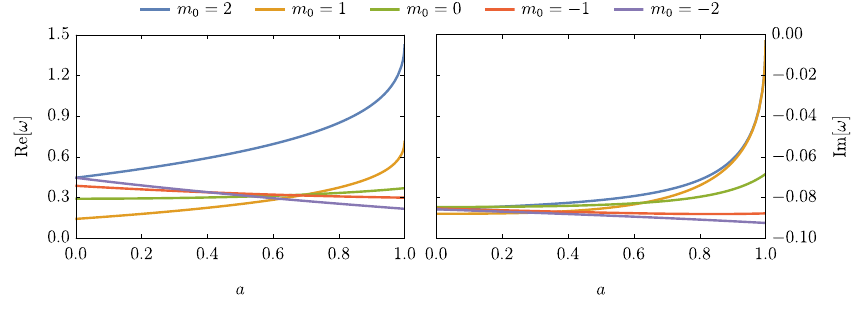}
	\caption{
		Real (left) and imaginary (right) parts of $ n=0 $ PS modes for $ s=-2$, $l=2$, $m_0 = 2, 1, 0, -1, -2$ with $A=0.2$.	
	}
	\label{fig:psac02}
\end{figure*}

\begin{figure*}[htbp]
	\centering
	\includegraphics[width=1\textwidth]{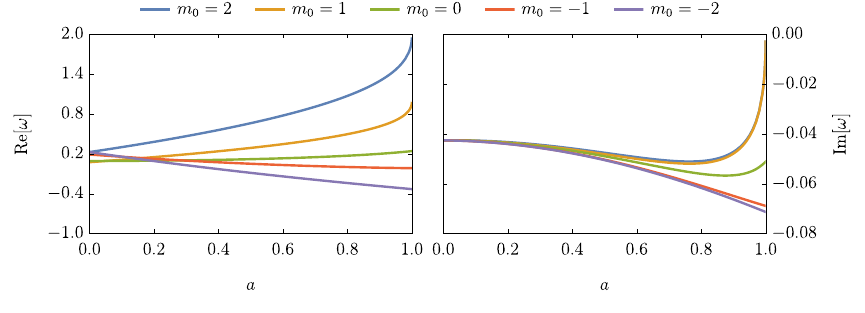}
	\caption{
            Real (left) and imaginary (right) parts of $ n=0 $ PS modes for $ s=-2$, $l=2$, $m_0 = 2, 1, 0, -1, -2$ with $A=0.4$.	
	}
	\label{fig:psac04}
\end{figure*}

\subsection{Photon sphere modes}
There is a well-known geometric correspondence between high-frequency quasinormal modes of black holes and properties of null geodesics that reside on the photon sphere. 
In the eikonal limit ($ m \sim l $), the QNM of stationary, spherically symmetric black holes are directly related to the frequency and the Lyapunov exponent of the null geodesics near the photon sphere~\cite{Ferrari:1984zz, Cardoso:2008bp}. 
For Kerr black holes, the correspondence between the QNMs frequencies and the orbital and precessional frequencies of the spherical photon orbit has been shown in~\cite{Ferrari:1984zz, Yang:2012he}. 
All the QNMs we obtained in this section reduce to the PS modes for Kerr black holes in the limit $ A \rightarrow 0 $ and we call them PS modes too.

We calculate the fundamental PS modes' frequencies for the gravitational perturbations with $ l=2 $. 
There are five different modes $ m_0 = \pm 2, \pm 1, 0 $ for the angular number $ l=2 $.
%since the azimuthal number $ m $ satisfies $ |m| \leq l $. 
The influences of spin and acceleration are both investigated. 
We first set $ a $ fixed and analyze the influences of the acceleration parameter $A$. 
As mentioned in Section \ref{sec:bkmetric}, the max value of the acceleration parameter $ A $ is $ \frac{1}{r_+} $. 
The results are shown in Figs.~\ref{fig:psa0}-\ref{fig:psa09}. 
The acceleration parameter has distinct influences on the quasinormal modes with different $m_0$, especially for the $m_0 = 1$ modes. 
Then we set $ A $ fixed and analyze the influences of the spin parameter $ a $. 
We show the results in Figs.~\ref{fig:psac0}-\ref{fig:psac04}. 
The QNMs we obtain in this section return to the QNMs of Kerr black holes when $ A \rightarrow 0 $.  

\begin{figure*}[htbp]
	\centering
	\includegraphics[width=1\textwidth]{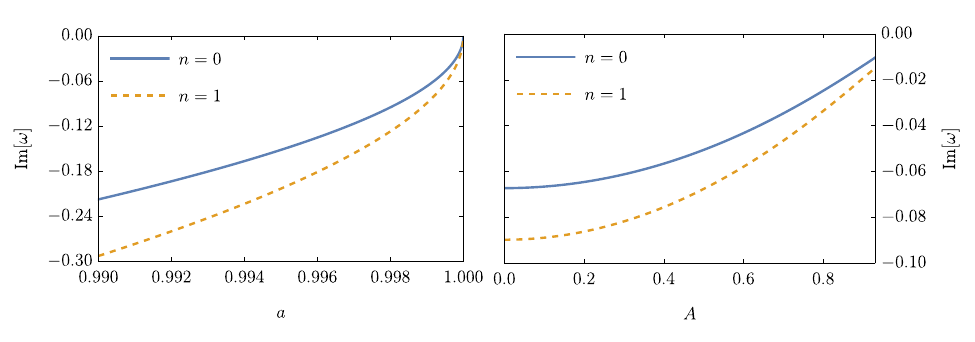}
	\caption{
            Imaginary parts of $ n = 0 $ and $ 1 $ NE modes for $ s=-2$, $l=2$, $m_0 =0$. 
            The left panel shows $ \text{Im}(\omega_\text{NE}) $ as a function of $ a $ with $A = 0.05$. 
            The right panel shows $ \text{Im}(\omega_\text{NE}) $ as a function of $ A $ with $a = 0.999$.	
            We can only reach $ A =0.93 $ during our numerical computation. 
            The extremal value of $ A $ is $ 0.957$. 	
	}
	\label{fig:ne}
\end{figure*}

Fig.~\ref{fig:psa0} shows the frequencies for $ a=0 $, which stands for the accelerating Schwarzschild black hole. 
When $ a=0 $, the angular perturbation equation (\ref{eq:angeq}) becomes independent of $ \omega $. 
The separation constant becomes a real number and can be obtained by solving the angular perturbation equation directly. 
When $ A \rightarrow 0 $, the C-metric becomes the Schwarzschild metric. 
The QNMs approach the Schwarzschild case and modes with different $ m_0 $ become degenerate. 
The black points in Fig.~\ref{fig:psa0} represent the $ n=0 $ fundamental mode of the Schwarzschild black hole. 
When $ A $ increases, the real parts of $ m_0 = 2, -1, -2$ modes increase to a maximum and then decrease, while the real part of $ m_0 = 0$ mode monotonically decreases. 
All the imaginary parts of $ m_0 = 2, 0, -1, -2$ modes increase with $ A $ increasing. 
As $ a=0 $, the frequencies have an additional symmetry -- they are symmetric about the imaginary axis. 
From the symmetry of the perturbation equations and the boundary conditions, it is obvious that the  $ -\omega^* $ and $ A_{lm} $ are also the eigenvalues if $ \omega $ and $ A_{lm} $ are eigenvalues for the mode having the azimuthal number $m$. 
Therefore, the $ m_0 = \pm 2$ modes coincide. 
The QNM of $ m_0 = 1 $ has an abnormal behavior. 
This is due to the conical singularity.
The deficit angle changes with the increment of $A$.
Thus, the boundary condition (\ref{eq:abdm}) is not smooth and $ s + m_0 \frac{P(\pi)}{P(0)} $ changes signs at $ \frac{P(\pi)}{P(0)} = 2 $ for $ s=-2, m_0 =1 $. 
This makes the $ m_0 = 1 $ mode very different from other modes and have a turning point at $ \frac{P(\pi)}{P(0)} = 2 $. 
We can also see that the $ m_0 = 1 $ mode has the largest imaginary part when $ A $ is small. 

In Figs.~\ref{fig:psa05}-\ref{fig:psa09}, we show the results for larger $ a $. 
When $ A \rightarrow 0 $, the QNM frequencies coincide with the Kerr results. 
The no-zero $ a $ results in the variance of frequencies for different $ m_0 $. 
For $ a \neq 0 $, the dominant mode changes, and $ m_0 = 2 $ mode becomes the dominant one when $ a $ is large enough.

When $ A $ approaches the extreme value (Nariai-type extremal condition), all PS modes' imaginary parts tend to $0$, but their real parts tend to a finite value around $ m \Omega_H(r_+) $. 
The real parts of $ m_0 = -1, -2 $ modes change their signs during this procedure. 
A similar phenomenon appears when we increase the spin parameter $a$ as shown in Fig.~\ref{fig:psac04}.
These are due to a dragging effect from the acceleration and rotation.
No fundamental quasinormal mode of the Kerr black hole changes the sign of its real part as $a$ increases. So does the fundamental mode of the C-metric when $A$ increases, which is shown in Fig.~\ref{fig:psa0}.
This phenomenon has also been observed for the Kerr-de Sitter case~\cite{Yoshida:2010zzb}. 
In the Nariai-type extremal limit, the imaginary parts of the PS modes can be approximated by
\begin{equation}
	\text{Im}(\omega_\text{PS}) \simeq  - (n + 1/2) [\kappa(r_+)+\kappa(r_A)]/2,
\end{equation}
which is consistent with the analytic approximation of the scalar perturbations~\cite{Gwak:2022nsi}.

Figs.~\ref{fig:psac0}-\ref{fig:psac04} are results with $ A $ fixed. 
In Fig.~\ref{fig:psac0}, the acceleration parameter $ A $ is set to be $ 10^{-4} $. 
Therefore, the influence of the acceleration is very small and the results are almost the same as the Kerr case. 
From these figures, we can see that the increment of spin parameter $ a $ tends to increase the real parts of positive modes ($ m_0 = 2, 1, 0 $) and decreases the real parts of negative modes ($ m_0 = -1, -2$). 
%The QNMs of Kerr black holes have similar results. 
The influence of $ a $ on the imaginary parts is more complex and highly dependent on the parameter $ A $. 
When $ A $ is large, as shown in Fig.~\ref{fig:psac04}, the imaginary parts of positive modes decrease first and then increase, while the negative modes' imaginary parts keep decreasing. 
The $ m_0 = \pm 2 $ modes have the same frequencies at $ a=0 $ because of the symmetry. Similarly, the $ m_0 = 1$ modes show an anomalous behavior.

\subsection{Near-extremal modes}
The spectrum of quasinormal modes bifurcates and a new distinct set of modes arises when the Cauchy and event horizon approach each other. 
This phenomenon has been found in the spectrum of nearly extremal Kerr BHs for some specific $ (l,m) $ pairs~\cite{PhysRevD.87.041502, Yang:2013uba, Richartz:2015saa}. 
For the Kerr black hole, one set of modes has a vanishing imaginary part while the other set's imaginary part remains a finite value in the extremal limit $ a \rightarrow 1 $.
They are called the zero-damping and damped modes in \cite{Yang:2013uba}. 
For the spinning C-metric, we focus on the $ m_0 = 0 $ modes, whose zero-damping modes and damped modes are very distinguishable. 
The fundamental PS modes for $ m_0 = 0 $ are all damped, as shown in Figs.~\ref{fig:psac0}-\ref{fig:psac04}. 
The zero-damping modes appear when $ a $ increases, and are called the near-extremal modes here.

\begin{figure*}[htbp]
	\centering
	\includegraphics[width=1\textwidth]{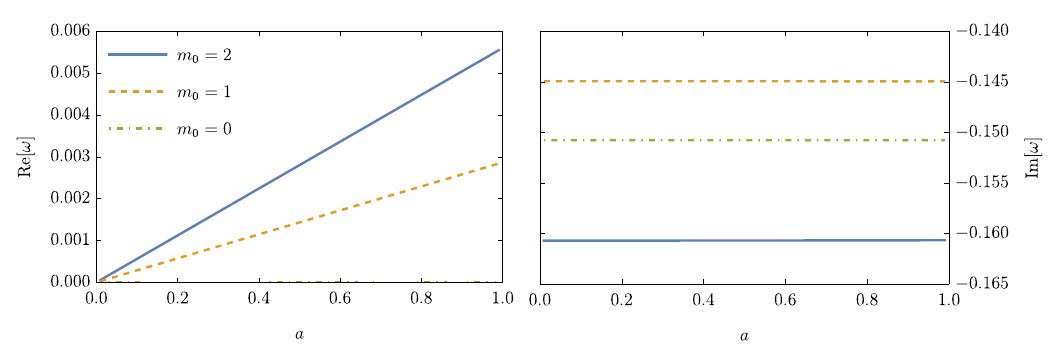}
	\caption{
            Real (left) and imaginary (right) parts of $ n=0 $ A modes for $ s=-2$, $l=2$, $m_0=2, 1, 0 $ with $A=0.05$.			
	}
	\label{fig:s2ac1}
\end{figure*}

\begin{figure*}[htbp]
	\centering
	\includegraphics[width=1\textwidth]{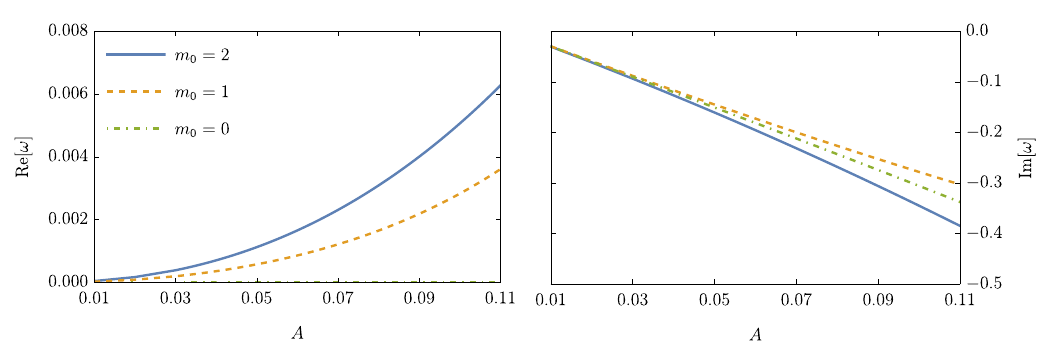}
	\caption{
            Real (left) and imaginary (right) parts of $ n=0 $ A modes for $ s=-2$, $l=2$, $m_0=2, 1, 0$ with $a=0.2$.			
	}
	\label{fig:s2ac2}
\end{figure*}

The near-extremal modes for $ m_0 = 0 $ have vanishing real parts. Fig.~\ref{fig:ne} shows the imaginary parts of the first two modes of this new family of modes. 
We can see that these modes' imaginary parts truly go to $ 0 $ from the left panel of Fig.~\ref{fig:ne}. At the limit $ a \rightarrow 1 $, we can approximate these modes by
\begin{equation}
	\omega_\text{NE} \simeq  -i [ \kappa(r_+)-\kappa(r_-)](n+ l +1)/2.
\end{equation}
In the right panel of Fig.~\ref{fig:ne}, we show the frequencies of the NE modes with the spin parameter $ a $ fixed. 
We can see that the imaginary parts also seem to approach zero with the acceleration parameter $ A $ increasing. 
However, when $A$ approaches the extremal value, the PS modes also approach zero as shown in the previous figures.
The numerical stability of the NE modes becomes much weaker. 
We can only calculate the results till $ A = 0.93 $. 
Similar modes have also been found in the spectrum of RN de Sitter BHs and charged accelerating BHs~\cite{Cardoso:2017soq, Destounis:2020pjk}.

%\begin{figure}[htbp]
%	\centering
%	\includegraphics[width=0.35\textwidth]{s2ne2.pdf}
%	\caption{
%		Imaginary parts of $ n = 0 $ and $ 1 $ NE modes for $ s=-2$, $l=2$, $m_0 =0$ with $a = 0.999$.			
%	}
%	\label{fig:ne2}
%\end{figure}

\subsection{Acceleration mode}
Acceleration modes are quasinormal modes that are highly dependent on the acceleration parameter. 
Their appearance is due to the acceleration horizon of the spinning C-metric.
Acceleration modes' imaginary parts have a linear dependence on $A$ and a very weak dependence on $a$.
These modes have been found in the scalar spectrum of charged or spinning C-metric~\cite{Destounis:2020pjk, Xiong:2023usm} and share similarities with the de Sitter modes for the RN-dS BHs~\cite{Cardoso:2017soq}. 
For the gravitational perturbations, we also identify these modes. The numerical results are shown in Figs.~\ref{fig:s2ac1}-\ref{fig:s2ac2}. 
We only show the positive modes for simplicity.

In Fig.~\ref{fig:s2ac1}, we fix the acceleration parameter $ A $ to be $ 0.05 $ and study the influences of the spin parameter $ a $. 
We can see that the imaginary parts are almost independent of $ a $ and are all around $ -0.15$. 
The $ m_0 = 1 $ mode decays most slowly, probably because of its special behavior under the angular boundary conditions. 
When $ m_0 \neq 0$, the real parts of the acceleration modes get small nonzero values. They are almost linearly dependent on $ a $. 
%The $ m_0 = 0 $ mode has a vanishing real part.

Fig.~\ref{fig:s2ac2} shows the results with $ a $ fixed. When $A$ increases, the $ m_0 = 0$ mode still has vanishing real parts, while the real parts of $ m_0 = 1, 2 $ modes increase. 
It is obvious that their imaginary parts are linearly dependent on the acceleration parameter $A$, which is also the surface gravity $\kappa_A^\text{R}$ at the acceleration horizon of Rindler space. 
Similarly, the $ m_0 = 1 $ mode is the dominant one. 
The imaginary parts of acceleration modes with $ m_0 \ge 0, m_0 \neq 1 $ can be approximated by
\begin{equation}
	\text{Im}(\omega_A) \simeq -  \kappa_A^\text{R} [ n + l + 1 + m_0(P(\pi)-1)]
\end{equation}
for small $ A $. The imaginary parts of $ m_0 = 1 $ modes can be approximated by 
\begin{equation}
	\text{Im}(\omega_A) \simeq -  \kappa_A^\text{R} [ n + l + P(0)]
\end{equation}
for small $ A $.

Compared with the other two sets, the numerical stability of this family is much weaker. 
We fail to find these modes when $ A $ is too small or too large.

\section{\label{sec:conclu} Conclusion and Discussion}
In this paper, we focused on the gravitational quasinormal modes of the spinning C-metric.
We showed that the perturbation equations for the spinning C-metric for massless perturbations with any spin weights can be transformed into the Heun's equation. 
The continued fraction method and shooting method were used to obtain the separation constant and the quasinormal mode frequency. 
The transformation of the equation makes the numerical calculation quicker and more precise. 

We identified three distinct sets of quasinormal modes. 
The influences of the spin and acceleration on the quasinormal frequencies were analyzed, including the extremal cases. 
The first set is the photon sphere modes, which dominate against the other modes for most of the parameter space. 
The acceleration parameter has very distinct influences on the quasinormal modes with different $m_0$.
With the increase of $A$, the real parts of the QNM frequencies with negative azimuthal numbers change signs. 
We also found that the $ s=-2, l=2, m_0=1 $ modes have an anomalous behavior. 
This is because the acceleration changes the angular boundary conditions. 
When the acceleration parameter approaches to its extremal value, the imaginary parts of these modes become vanishing. 
The second set is the near-extremal modes, which branch from the first set when the spinning parameter $ a $ increases. 
This family of modes becomes dominant in the extremal limit $ a \rightarrow 1 $. 
The last set is the acceleration modes. 
They are closely related to the acceleration parameter and decay most slowly when $ A $ is small. 
Empirical formulas were given to approximate these modes in the low acceleration or extremal limits. 

There are still some unsolved problems with the QNM spectrum of the spinning C-metric, such as a more detailed analysis of the spectrum of the near-extremal spinning C-metric or the spectrum of the electromagnetic perturbation. 
Although we only analyzed the gravitational quasinormal modes of the spinning C-metric, our methods can calculate QNMs of all the perturbations of the spinning C-metric (including the C-metric).

The C-metric has been used to approximate the accelerating black hole in~\cite{PhysRevD.107.044031, PhysRevLett.129.031102}. 
Here we extended the previous works and analyzed the gravitational QNM spectrum of the spinning C-metric. 
From our numerical results, we can see that small acceleration can still have a relatively large effect on the QNM frequencies of spinning C-metric. 
If an accelerating rotating black hole can be described by the spinning C-metric, this phenomenon can be used to detect the acceleration of the black hole. 
However, there are still some problems with the spinning C-metric. 
First, the C-metric is not strictly asymptotically flat. 
The generators of its null infinity are not complete~\cite{Ashtekar:1981ar}.
In addition, the C-metric has two conical singularities and one of them cannot be removed. 
This is the sacrifice of accelerating the black hole without the introduction of external fields.
How to precisely define gravitational waves in such spacetime is still an open question.
%Moreover, the stationary area of the spinning C-metric is between the event horizon and the acceleration horizon. In this area, the curvature remains finite. If we want to use it to approximate the accelerating black hole in the astrophysical environment, we should set the acceleration to be very small.
%The different ways of removing the conical singularity will result in different QNM frequencies. 
We need further study to have a better understanding of the spinning C-metric. 

\begin{acknowledgments}
This work is supported in part by 
the National Key Research and Development Program of China Grant  No. 2021YFC2203004, No. 2020YFC2201502 and No. 2021YFA0718304,
the National Natural Science Foundation of China Grants No. 12105344, No. 11821505, No. 11991052, No. 11947302, and No. 12047503, No.12235019 and the Science Research Grants from the China Manned Space Project with No. CMS-CSST-2021-B01.
\end{acknowledgments}

\appendix

\section{\label{app:NPquantity}Newman-Penrose formalism}
The NP formalism is a special case of the tetrad formalism, where the tensors of the theory are projected onto a vector basis.
%A specific choice of vector basis can lead to simplified expressions. 
Here we introduce the definitions of the quantities used in Section \ref{sec:perturbME}. 
For more details, interested readers may refer to \cite{Newman:1961qr, Teukolsky:1973ha, 10.1093/oso/9780198503705.001.0001}.

Assume that we choose a null tetrad $\{l, n, m, \bar{m} \}$. For the signature $(-, +, +, +)$, the normalization convention is
\begin{equation}
	l^a n_a = -1, \quad m^a \bar{m}_a =1,
\end{equation}
while all other scalar products are zero.
The primary quantities used in the NP formalism are twelve spin coefficients, five Weyl-NP scalars, and ten Ricci-NP scalars. Since we consider the vacuum case, all Ricci-NP scalars vanish and we do not show their definitions here. The spin coefficients are defined by
\begin{eqnarray}
    \kappa &\equiv& - m^a l^b \nabla_b l_a, \qquad \tau = - m^a n^b \nabla_b l_a, \nonumber \\
    \sigma &\equiv& - m^a m^b \nabla_b l_a, \qquad \rho \equiv -m^a \bar{m}^b \nabla_b l_a, \nonumber \\
    \pi &\equiv& \bar{m}^a l^b \nabla_b n_a, \qquad \nu \equiv \bar{m}^a n^b \nabla_b n_a, \nonumber \\
    \mu &\equiv& \bar{m}^a m^b \nabla_b n_a, \qquad \lambda \equiv \bar{m}^a \bar{m}^b \nabla_b n_a, \nonumber \\
    \epsilon &\equiv& -\frac{1}{2} (n^a l^b \nabla_b l_a - \bar{m}^a l^b \nabla_b m_a), \nonumber \\
    \gamma &\equiv& -\frac{1}{2} (n^a n^b \nabla_b l_a - \bar{m}^a n^b \nabla_b m_a), \nonumber \\
    \beta &\equiv& -\frac{1}{2} (n^a m^b \nabla_b l_a - \bar{m}^a m^b \nabla_b m_a), \nonumber \\
    \alpha &\equiv& -\frac{1}{2} (n^a \bar{m}^b \nabla_b l_a - \bar{m}^a \bar{m}^b \nabla_b m_a).
\end{eqnarray}
The Weyl-NP scalars are defined by
\begin{eqnarray}
    \Psi_0 &\equiv& C_{abcd} l^a m^b l^c m^d, \qquad \Psi_1 \equiv C_{abcd} l^a n^b l^c m^d, \nonumber \\
    \Psi_2 &\equiv& C_{abcd} l^a m^b \bar{m}^c n^d, \qquad \Psi_3 \equiv C_{abcd} l^a n^b \bar{m}^c n^d, \nonumber \\
    \Psi_4 &\equiv& C_{abcd} n^a \bar{m}^b n^c \bar{m}^d.
\end{eqnarray}
			
In many situations, like the Petrov type D vacuum spacetimes \cite{10.1093/oso/9780198503705.001.0001}, the Newman-Penrose formalism simplifies dramatically. Many of the quantities vanish when we choose some specific null tetrad. This simplification makes it easier to do calculations than using the standard form of Einstein's equations.

\begin{table*}
	\caption{\label{tab:2} 
        The comparison between the gravitational quasinormal modes computed by two methods with different parameters. 
        We only present the fundamental mode with $ l=2 $ for each set. 
        In some extremal cases, the shooting method fails to get the results.}
	\begin{ruledtabular}
		\begin{tabular}{ccccc}
			Parameters & Families & Azimuthal number & Continued fraction method & shooting method \\[0.6ex] \hline
			&   & $ m_0 = 2 $ & $ 0.4823156711 - 0.0892117509 i $ &  $ 0.4823156711 - 0.0892117509 i $  \\
			$ a = 0.2 $ &    & $ m_0 = 1 $ & $ 0.3026513887 - 0.0867337232 i $ &  $ 0.3026513887 - 0.0867337232 i $  \\
			$ A = 0.1 $ & PS & $ m_0 = 0 $ & $ 0.3534844386 - 0.0885607886 i $ &  $ 0.3534844386 - 0.0885607886 i $  \\
			&    & $ m_0 = -1 $ & $ 0.3862871232 - 0.0900322898 i $ &  $ 0.3862871232 - 0.0900322898 i$  \\
			&    & $ m_0 = -2 $ & $ 0.4082687782 - 0.0909585115 i $ &  $ 0.4082687782 - 0.0909585115 i$  \\[0.6ex]
			$ a = 0.2 $ &   & $ m_0 = 2 $ & $ 0.2037690459 - 0.0000502429 i $ &  $\cdots$  \\
			$ A = 0.505 $ &  PS  & $ m_0 = 1 $ & $ 0.1019309535 - 0.0000502427 i $ &  $ \cdots $  \\
			&   & $ m_0 = 0 $ & $ 0.0000502527 - 0.0000502422 i $ &  $ \cdots $   \\[0.6ex]
			% $ a = 0.2 $ &    & $ m_0 = 2 $ & $ 0.0011204094 - 0.1606971968 i $ &  $ 0.0011204094 - 0.1606971968 i$  \\
			% $ A = 0.05 $ &  A  & $ m_0 = 1 $ & $ 0.0005730740 - 0.1449438445 i $ &  $ 0.0005730740 - 0.1449438445 i$  \\
			% &     & $ m_0 = 0 $ & $ -0.1507749493 i $ &  $ -0.1507749493 i$  \\[0.6ex]
			$ a = 0.2 $ &    & $ m_0 = 2 $ & $ 0.0000408386 - 0.0304060717 i $ &  $ 0.0000408386 - 0.0304060717 i $ \\
			$ A = 0.01 $ &  A  & $ m_0 = 1 $ & $ 0.0000204331 - 0.0298001931 i $ & $ 0.0000204331 - 0.0298001931 i $  \\
			&     & $ m_0 = 0 $ & $ -0.0300063368 i $ &  $ -0.0300063368 i $   \\[0.6ex]
			$ a=0.999, A = 0.1 $ &  NE  & $ m_0 = 0 $ & $ -0.0666019203 i $ &  $ -0.0666019203 i$  \\[0.6ex]
			$ a=0.999999, A = 0.1 $ &  NE  & $ m_0 = 0 $ & $ -0.0021160230 i $ &  $ -0.0021160230 i$  \\
			%	$ -5.423853784 - 0.125559768 I $& $ -5.423853784 - 0.125559768 I $
		\end{tabular}
	\end{ruledtabular}
\end{table*}

\begin{table*}
	\caption{\label{tab:3} 
        Numerical results for $ s= -2 $ QNMs with $ A = 10^{-10} $ or for $ s = 0 $ QNMs. 
        The results computed by both methods are all the same except for the $ A = 10^{-10} $ cases, where the shooting method fails. 
        We stop comparing the results from the two methods for brevity. 
        The subscript stands for the overtone number. The $ s= -2 $ QNMs are well consistent with the Kerr case. 
        The scalar QNMs we show here were calculated in \cite{Destounis:2020pjk, Xiong:2023usm}. 
        Our results agree with them.}
	\begin{ruledtabular}
		\begin{tabular}{ccccc}
			Parameters & Families & Azimuthal number & QNM frequencies & Separation constant \\[0.6ex] \hline
			$ s=-2 $&   & $ m_0 = 2 $ & $ 0.3870175385 - 0.0887056990 i $ &  $ -3.742360877 - 0.058957785 i $  \\
			$ l=2 $ &    & $ m_0 = 1 $ & $ 0.3804322549 - 0.0887983009 i $ &  $ -3.873766302 - 0.029315246 i $  \\
			$ a = 0.1 $ & $\text{PS}_0$ & $ m_0 = 0 $ & $ 0.3740317881 - 0.0888980902 i $ &  $-4.000628538 + 0.000316649 i $  \\
			 $ A = 10^{-10} $ &   & $ m_0 = -1 $ & $ 0.3678117862 - 0.0890036975 i $ &  $  -4.123144955 + 0.029946687 i$  \\
			&    & $ m_0 = -2 $ & $ 0.3617677061 - 0.0891137673 i $ &  $ -4.241505977 + 0.059581772 i$  \\[0.6ex]
			$ s=0, l=1 $ & $\text{PS}_0$ & $ m_0 = 1 $ & $ 0.3032499295 - 0.0973670415 i $ &  $-2.184342759 $  \\
			$ a = 0, A = 0.03 $ &  $\text{A}_0$  &   & $ -0.0618960744 i $ &  $ -2.184342759 $  \\[0.6ex]
			$ s=0,l=0 $ &  $ \text{NE}_0 $  &   & $ -0.0002230478576 i $ &  $ 3.358418504*10^{-8}$  \\
			$ a=1-10^{-7} $ & $ \text{NE}_1 $  & $ m_0 = 0 $ & $ -0.0004460958815 i $ &  $ 1.333343335*10^{-7} $  \\
			$ A = 0.05 $& $ \text{NE}_2 $ &   & $ -0.0006691440726 i $ &  $ 2.995847651*10^{-7}$  \\[0.6ex]
			$ s=0, l=0$ & $ \text{A}_0 $ &   & $ -0.002000027032 i $ &  $ 2.666696125*10^{-6}$  \\
			$ a = 0.5 $ & $ \text{A}_1 $ & $ m_0 = 0 $ & $ -0.004000173569 i $ &  $ 4.666918426*10^{-6}$  \\
			$ A=0.002 $& $ \text{A}_2 $ &   & $ -0.006000441735 i $ &  $ 8.000919614*10^{-6} $  \\[0.6ex]
			$ s=0, l = 0 $ &  $\text{PS}_0$  &   & $ 0.002228495959 - 0.001796566429 i $ &  $ -0.2596500358 + 7.026*10^{-7} i$  \\
			$ a=0.3 $ &  $\text{PS}_1$  & $ m_0 = 0 $ & $ 0.001049501306 - 0.005389700536 i $ &  $ 0.3217182425 + 1.4275*10^{-6} i$  \\
			$ A=0.508 $ & $\text{PS}_2$ &    & $ 0.001049338537 - 0.008982834229 i $ &  $ 0.3217247589 + 2.3788*10^{-6} i$  \\
			%	$ -5.423853784 - 0.125559768 I $& $ -5.423853784 - 0.125559768 I $
		\end{tabular}
	\end{ruledtabular}
\end{table*}

\section{\label{app:coefficients}Coefficients in the equations}
In this appendix, we show the coefficients in Eq.~(\ref{eq:asimp}):
\begin{eqnarray}
		\frac{\dif^2 S(z)}{\dif z^2} &+& \left(\frac{1}{z} + \frac{1}{z-1} + \frac{1}{z-z_+} - \frac{2}{z-z_\infty} \right) \frac{\dif S(z)}{\dif z } \nonumber \\
		&+&\bigg[ -\frac{A_1^2}{z^2} -\frac{A_2^2}{(z-1)^2} -\frac{A_3^2}{(z-z_+)^2}  \nonumber \\
		&+&\frac{2}{(z-z_\infty)^2} + \frac{A'_1}{z} + \frac{A'_2}{z-1} + \frac{A'_3}{z-z_+}  \nonumber \\
		&-& \frac{\frac{1}{z_\infty} + \frac{1}{z_\infty -1 } + \frac{1}{z_\infty - z_+}}{z-z_\infty} \bigg] S(z).
\end{eqnarray}
The coefficients $A'_1, A'_2, A'_3$ are
\begin{eqnarray}
	A'_1 &=& -\frac{A_{lm}}{K_+} - \frac{2(\ta^2 + \tA)K_-}{P(0) P(\pi)} \nonumber \\
	& &+ \frac{m^2 [-1 + 3 \ta^4 + 4 \tA(-1 + \tA + 2 L )]}{2P(0)P(\pi)^3} \nonumber \\
	& &+ \frac{2 m^2 \ta^2 (-3 + 2 \tA + 4 L)}{2P(0)P(\pi)^3} \nonumber \\
	& &+ \frac{2m(\ta^2 - \tA + L) + (1-\ta^2) K_-}{P(0) P(\pi)} s \nonumber \\
	& &+ \frac{1- 4L + 4 \tA^2 + \ta^2(-6 + \ta^2 + 4L)}{2P(0) P(\pi)} s^2 \nonumber \\
	& &+ 2 \tomega \bigg[ \frac{m K_-}{P(0) P(\pi)^2} + \frac{s}{K_+} \bigg],
\end{eqnarray}

\begin{eqnarray}
	A'_2 &=& \frac{A_{lm}}{K_-} + \frac{2(\ta^2 - \tA)K_+}{P(0) P(\pi)} \nonumber \\
	& &+ \frac{m^2 [1 - 3 \ta^4 - 4 \tA(1 + \tA + 2 L )]}{2P(0)^3 P(\pi)} \nonumber \\
	& &+ \frac{2 m^2 \ta^2 (3 + 2 \tA + 4 L)}{2P(0)^3 P(\pi)} \nonumber \\
	& &+ \frac{2m(\ta^2 + \tA - L) - (1-\ta^2) K_+}{P(0) P(\pi)} s \nonumber \\
	& &- \frac{1+ 4L + 4 \tA^2 + \ta^2(-6 + \ta^2 - 4L)}{2P(0) P(\pi)} s^2 \nonumber \\
	& &+ 2 \tomega \bigg[ -\frac{m K_+}{P(0)^2 P(\pi)} + \frac{s}{K_-} \bigg],
\end{eqnarray}
	
\begin{eqnarray}
	A'_3 &=& - \frac{4 L A_{lm}}{P(0) P(\pi)} + \frac{8 L^2 (\tA - L)}{\ta^2 P(0) P(\pi)} \nonumber \\
	& &- \frac{8 m^2}{[P(0) P(\pi)]^3} \bigg[ \ta^6 (-\tA + L) + 2 \ta^4 (\tA + L) \nonumber \\
	& &- 4 \tA^2 (\tA + L) + \ta^2 (3 \tA + L) \bigg] \nonumber \\
	& &+ \frac{4 L (1-\ta^2)}{P(0) P(\pi)} s^2 - \frac{4(m \ta^2 - L(1-\ta^2))}{P(0)P(\pi)}s \nonumber \\
	& &+ \frac{8 m \tomega (L + \tA -\tA \ta^2 + L \ta^2)}{[P(0) P(\pi)]^2}  \nonumber \\
	& &-\frac{4 (1-\ta^2) s \tomega}{P(0) P(\pi)},
\end{eqnarray}
where
\begin{eqnarray}
	K_\pm = 1-\ta^2 \pm 2 L,\qquad L = \sqrt{\tA^2 -\ta^2}.
\end{eqnarray}

\section{\label{sec:commethod} Comparison of the two methods}
In this appendix, we show the numerical results for QNMs obtained by the continued fraction method and the shooting method. 

The comparison of the gravitational perturbation results computed by these two methods is shown in Table~\ref{tab:2}. 
We can see that they agree with each other very well. 
From our numerical calculations, we find that the continued fraction method is more fast and stable. 
The efficiency of the shooting method highly depends on the choice of the initial values. 
When the initial values depart from the true results too far, the time of computation increases dramatically. 
It also fails in some extremal cases as shown in Table~\ref{tab:2}.
In addition, we show some results for $ s = 0 $ scalar perturbations in Table~\ref{tab:3}. 
They are consistent with the results obtained in \cite{Destounis:2020pjk, Xiong:2023usm}. 
We also show the results for $ s = -2 $ gravitational perturbations with $ A = 10^{-10} $. 
The results agree with the Kerr QNMs very well~\cite{Berti:2009kk}.

\newpage

\nocite{*}

\bibliography{citeLib}% Produces the bibliography via BibTeX.

\end{document}